\documentclass[lettersize,journal]{IEEEtran}
\usepackage{amsmath,amsfonts}
\usepackage{algorithmic}
\usepackage{algorithm}
\usepackage{array}
\usepackage[caption=false,font=normalsize,labelfont=sf,textfont=sf]{subfig}
\usepackage{textcomp}
\usepackage{stfloats}
\usepackage{url}
\usepackage{verbatim}
\usepackage{graphicx}
\usepackage{cite}
\begin{document}

\title{WiFi Backscatter for Green Internet of Things: Concepts, Research Trends, and Practical Challenges}

\author{{Weiqi Wu, \textit{Member},~\textit{IEEE}, Shuo Wang, and Zhaoyuan Xu, \textit{Member},~\textit{IEEE}}
\thanks{Corresponding author: Shuo Wang.}
}



\maketitle

\begin{abstract}
WiFi backscatter has emerged as a promising technology for green Internet of Things (IoT) connectivity by enabling battery-free devices to communicate through widely available WiFi signals. Despite substantial progress in recent years, a considerable gap remains between research prototypes and practical deployment. This paper provides an overview of WiFi backscatter from the perspective of practical and green IoT systems. We first introduce the fundamentals and key enabling techniques, together with potential IoT applications. We then outline recent research trends toward higher throughput, concurrent communication, simplified deployment, commercial compatibility, and joint communication and sensing. Furthermore, we identify the key challenges that still hinder practical deployment, such as limited transmitter-to-tag operating range and packet loss in frequency-shifted backscatter. We believe that addressing these challenges will be critical to enabling WiFi backscatter to become a practical communication technology for future green IoT systems.
\end{abstract}


\section{Introduction}
The rapid growth of the Internet of Things (IoT) is enabling pervasive sensing, monitoring, and automation across a wide range of applications \cite{yuan2024native,du2023timespan}. At the same time, the increasing number of connected devices raises concerns about energy consumption, battery maintenance, and environmental sustainability. These challenges have driven growing interest in green IoT, which seeks to reduce device energy consumption and maintenance overhead while improving sustainability. Backscatter communication is particularly attractive for green IoT because a backscatter device, or tag, does not generate its own radio-frequency (RF) carrier \cite{kellogg2014wifi}. Instead, it conveys information by modulating and reflecting incident RF signals through changes in antenna impedance, which greatly reduces the energy required for wireless transmission. With energy harvesting, backscatter tags can operate with little or no battery dependence, enabling low-power and potentially battery-free IoT devices while reducing battery replacement and maintenance overhead. Ambient backscatter further reduces system overhead by reusing RF signals already available in the environment, eliminating the need for a dedicated carrier source \cite{zhang2016hitchhike}.

Among the available ambient RF sources, WiFi is especially appealing because of its widespread presence in homes, offices, hospitals, warehouses, and industrial environments \cite{wu2025concurrent,gong2025universal}. Early WiFi backscatter systems demonstrated that RF-powered tags could communicate with commodity WiFi infrastructure at data rates of up to $1$ Kbps over a distance of $2.1$ m, highlighting the potential of WiFi backscatter for low-power IoT connectivity \cite{kellogg2014wifi}. Nevertheless, practical deployment remains challenging. Backscattered signals are inherently weak and can be easily overwhelmed by strong direct-link WiFi transmissions and environmental interference. In addition, practical systems must provide compatibility with commodity WiFi devices and efficient support for multiple tags.

Subsequent research has attempted to address these challenges from several directions. FS-Backscatter \cite{zhang2016enabling} introduces frequency shifting to separate backscattered signals from strong ambient WiFi transmissions. HitchHike \cite{zhang2016hitchhike} improves compatibility by allowing commodity WiFi radios to recover tag information. Other studies have enabled multi-tag communication through concurrent signal separation and decoding \cite{wu2025concurrent,wang2024multirider}. These advances are moving WiFi backscatter toward practical deployment as an enabling technology for green IoT.

Given the growing role of WiFi backscatter in green IoT, this paper provides a comprehensive overview of WiFi backscatter from the perspective of practical green IoT deployment. The main contributions are summarized as follows:
\begin{itemize}
\item We introduce the fundamentals of WiFi backscatter and highlight its key enabling techniques and potential applications.
\item We summarize recent research trends toward higher throughput, concurrent communication, simplified deployment, commercial compatibility, and joint communication and sensing.
\item We identify the key challenges that still hinder practical deployment, including limited transmitter-to-tag operating range, packet loss in frequency-shifted backscatter, limited system generality, commodity-compatible concurrent communication, and the lack of standardization.
\end{itemize}

\section{WiFi Backscatter Overview}
\begin{figure*}
    \centering
    \includegraphics[width=\linewidth]{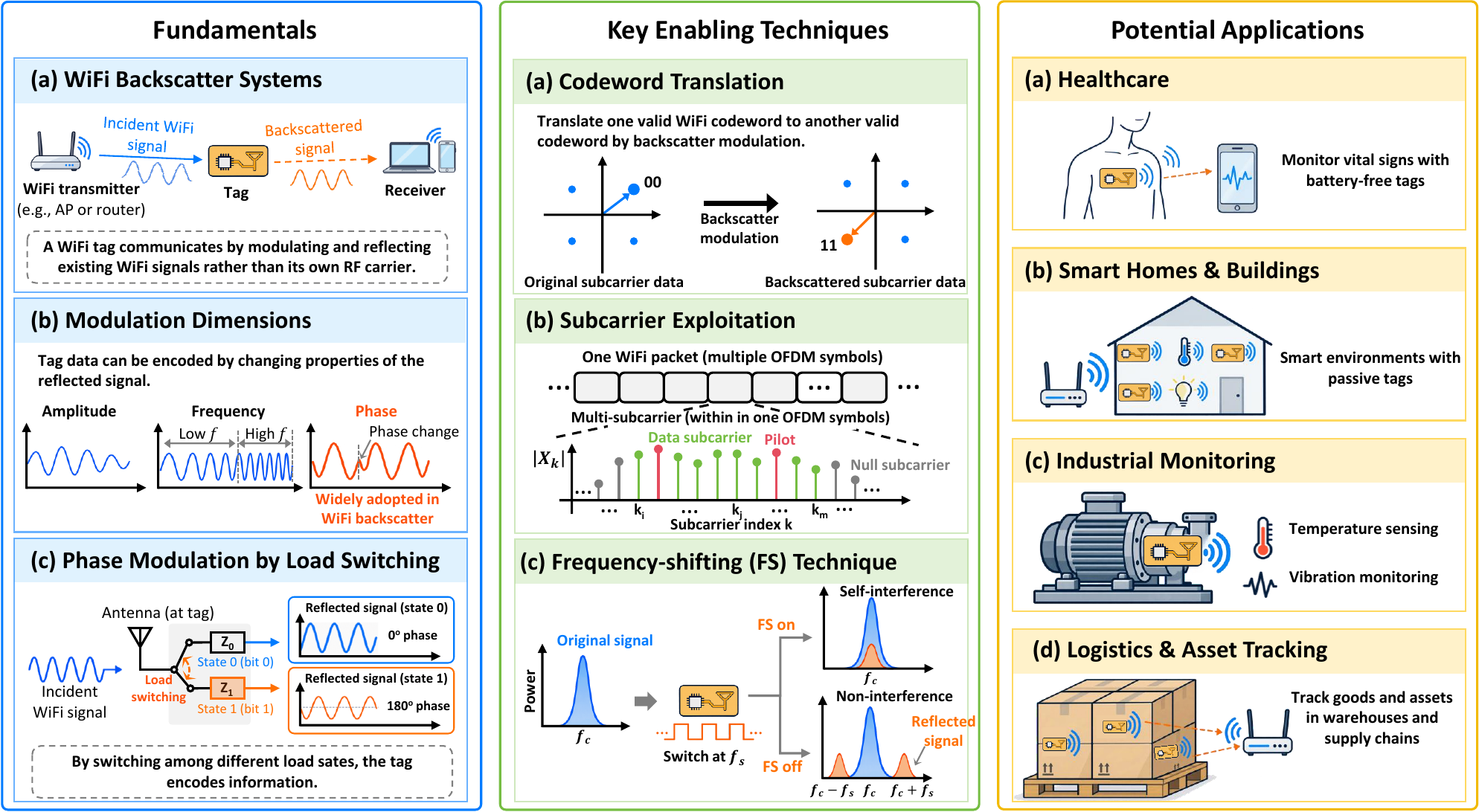}
    \caption{Overview of WiFi backscatter, including its fundamentals, key enabling techniques, and potential applications.}
    \label{fig_overview}
\end{figure*}

This section provides an overview of WiFi backscatter, covering its fundamentals, key enabling techniques, and potential applications, as outlined in Fig. \ref{fig_overview}.
\subsection{Fundamentals}

A WiFi backscatter tag communicates by modulating and reflecting existing WiFi signals rather than generating its own RF carrier. This principle is analogous to a mirror that conveys information by altering reflected light instead of producing light itself. A typical WiFi backscatter system consists of three components: a WiFi transmitter, one or more backscatter tags, and one or more receivers. The transmitter emits the 802.11 signal. The tags modulate the reflected waveform to embed their data, and the receiver extracts the tag information from the received signal.

Information can be encoded by modifying the amplitude, frequency, phase, or a combination of these properties of the reflected signal. Among these options, phase modulation is widely adopted in WiFi backscatter. In contrast, amplitude modulation is more sensitive to channel and signal-strength variations, while frequency-shift modulation requires the tag to periodically switch its reflection state to create a controlled frequency offset, typically introducing additional timing and control complexity. Phase modulation, by contrast, can be realized through simple switching between different antenna-load states.

To illustrate the tag modulation process, consider an incident WiFi signal $x(t)=s(t)e^{j2\pi f_ct}$, where $s(t)$ and $f_c$ denote the complex baseband waveform and carrier frequency, respectively. By switching among different load states, the tag changes its reflection coefficient $\Gamma_k=\rho_k e^{j\phi_k}$, where $\rho_k$ and $\phi_k$ denote the reflection magnitude and phase associated with state $k$. The resulting backscattered signal is $
x_b(t)=\Gamma_k x(t) =\rho_k s(t)e^{j(2\pi f_ct+\phi_k)}$. The tag can therefore encode information by selecting different reflection phases $\phi_k$. For instance, two load states with $\phi_k=0$ and $\phi_k=\pi$ introduce phase shifts of $0^\circ$ and $180^\circ$, respectively, allowing binary tag information to be represented by the phase of the backscattered WiFi signal.

\subsection{Key Enabling Techniques}
\textbf{Codeword Translation.} A wireless transmitter does not map arbitrary bit sequences to arbitrary waveforms. Instead, information is represented by a finite set of predefined signal patterns, which can be viewed as a \emph{codebook}, and each valid pattern is referred to as a \emph{codeword}. For example, 802.11b represents information using predefined spreading/code sequences, while orthogonal frequency-division multiplexing (OFDM)-based WiFi maps coded bits onto constellation points such as binary phase shift keying (BPSK) or quadrature-PSK (QPSK) symbols. The key idea of codeword translation is to transform one valid codeword into another valid codeword. Let $C=\{c_1,c_2,\ldots\}$ denote the set of valid codewords. A backscatter tag changes its reflection state such that an incident codeword $c_i$ is mapped to another $c_j\in C$ according to the tag bit. Because the backscattered signal still corresponds to a valid element of the original codebook, it can be decoded by a commodity WiFi receiver. Owing to its compatibility with commodity WiFi receivers, codeword translation has become a widely used technique in WiFi backscatter systems.

\textbf{Subcarrier Exploitation.}
Modern WiFi systems employ OFDM, in which each symbol is composed of multiple orthogonal subcarriers. For an \(N\)-point OFDM symbol, the transmitted time-domain signal can be expressed as $x[n]=\frac{1}{\sqrt{N}}\sum_{k=0}^{N-1}X_k e^{j2\pi kn/N}$, where $X_k$ denotes the frequency-domain symbol carried by the $k$-th subcarrier. These subcarriers have different functions: data subcarriers carry payload information, pilot subcarriers provide known reference symbols for channel estimation, and null subcarriers remain unused.

The structured nature of OFDM provides several opportunities for WiFi backscatter design. Some systems exploit information from individual subcarriers to improve tag-data recovery \cite{yu2023subscatter}. More importantly, the availability of multiple subcarriers within each OFDM symbol provides finer-grained frequency-domain resources for embedding tag data, allowing multiple tag symbols to be conveyed within a single WiFi packet and thereby substantially increasing backscatter throughput. Other systems leverage the known pilot subcarriers as inherent references for single-receiver demodulation \cite{wu2025concurrent}. In concurrent backscatter systems, differences across subcarriers can further provide additional dimensions for separating concurrent tag transmissions \cite{wang2024multirider,wu2026clusterfi}. Overall, exploiting the OFDM structure enables higher-throughput transmission, single-receiver decoding, and concurrent multi-tag communication.

\textbf{Frequency-shifting Technique.} Frequency-shifting \cite{zhang2016enabling} separates the weak backscattered signal from the strong direct-link WiFi signal in the frequency domain. Let the incident WiFi signal be $x(t)=s(t)e^{j2\pi f_c t}$. The tag switches its loads at a frequency $f_s$. This switching introduces frequency components around $f_c \pm f_s$, and the resulting backscattered signal can be expressed as $x_b(t)\propto s(t)e^{j2\pi (f_c\pm f_s)t}$. Thus, the reflected WiFi signal is shifted away from the original carrier and can be received on an adjacent non-overlapping WiFi channel.

\subsection{Potential Applications}
\textbf{Healthcare:} Wearable and implantable tags can leverage existing WiFi signals to support low-power and battery-free health monitoring. Such tags can be embedded in clothing, skin patches, or personal medical devices to continuously collect physiological data and transmit it to nearby commodity WiFi devices, including smartphones, laptops, and access points. By reducing or even eliminating the need for batteries, WiFi backscatter can extend device lifetime and lower the maintenance cost of long-term health monitoring systems.

\textbf{Smart Homes and Buildings:} The widespread availability of WiFi infrastructure makes homes and buildings particularly suitable for backscatter-enabled sensing. Battery-free tags can be integrated into doors, furniture, appliances, and building structures to monitor environmental conditions, human activities, and device status. By leveraging existing WiFi networks for both signal excitation and data reception, these tags can operate without dedicated readers or additional communication infrastructure.

\textbf{Industrial Monitoring:} Factories and other industrial environments often require continuous monitoring of machines, equipment, and infrastructure. Backscatter tags can be attached to machines, pipelines, tools, and structural components to monitor parameters such as vibration, temperature, operating conditions, and equipment status. Existing industrial WiFi infrastructure can then be reused for data communication, making such deployments well suited for long-term and low-maintenance monitoring.

\textbf{Logistics and Asset Tracking:} Warehouses and logistics systems provide another attractive setting for WiFi backscatter. Tags attached to packages, pallets, and equipment can use existing WiFi infrastructure for communication, reducing the need for dedicated RFID readers. Beyond identification and localization, these tags can also report sensing information such as temperature, humidity, and handling conditions, enabling integrated asset tracking and condition monitoring throughout the logistics process.

\section{Research Trends}
\begin{figure*}
    \centering
    \includegraphics[width=\linewidth]{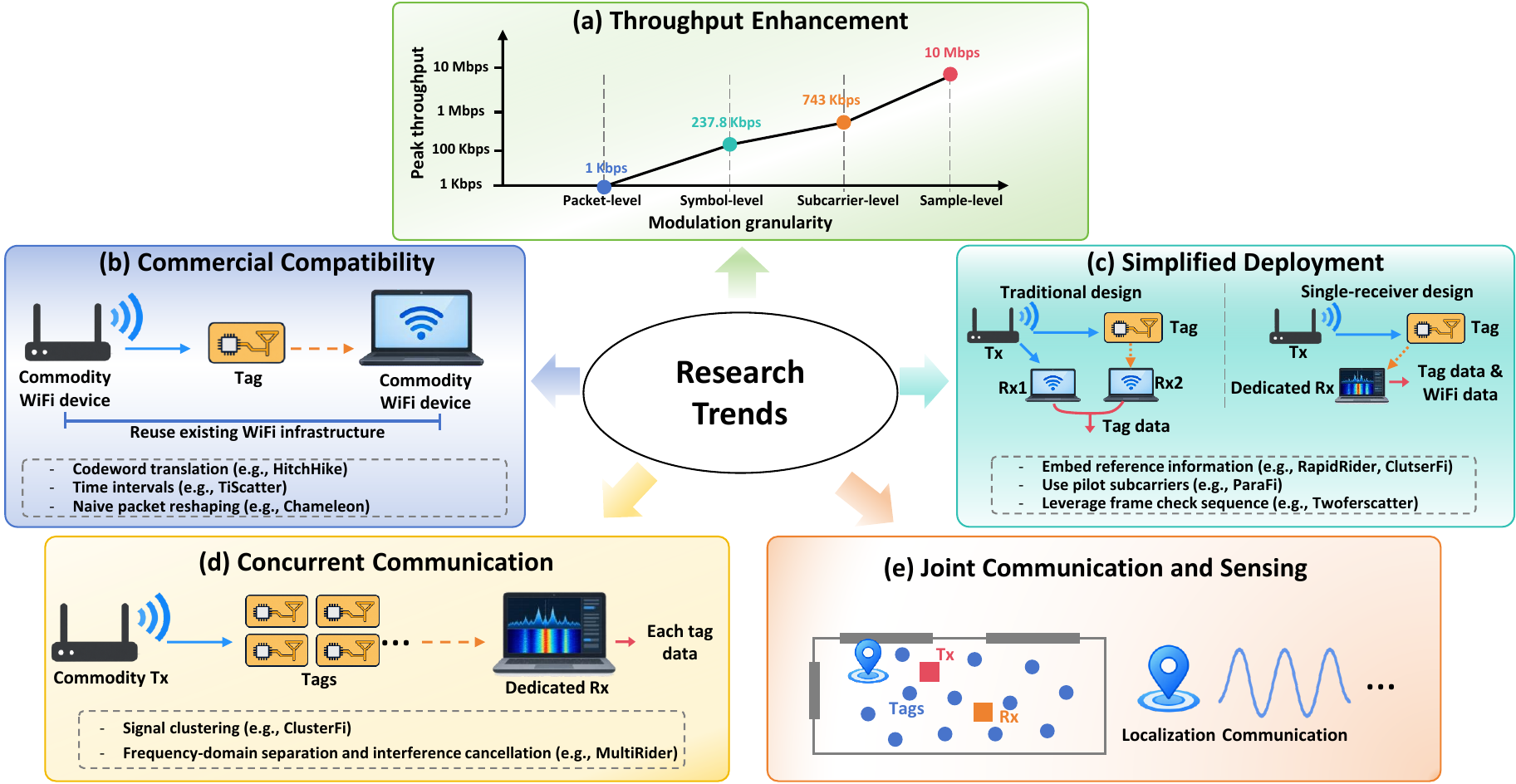}
    \caption{Research trends in WiFi backscatter toward practical green IoT, including throughput enhancement, commercial compatibility, simplified deployment, concurrent communication, and joint communication and sensing.}
    \label{fig_research_trends}
\end{figure*}

Recent advances in WiFi backscatter are opening new opportunities for practical green IoT applications. In this section, we highlight current research trends. Figure \ref{fig_research_trends} provides an overview.

\subsection{Throughput Enhancement}
Early WiFi backscatter systems operate at the packet level, where each packet carries only one or a few tag bits through coarse reflection-state control, achieving a peak throughput of $1$ Kbps \cite{kellogg2014wifi}. This coarse modulation granularity limits the amount of tag information that can be embedded into each WiFi packet, motivating subsequent designs to exploit finer-grained structures within the WiFi waveform. Symbol-level schemes encode tag data by manipulating individual OFDM symbols, increasing peak throughput to $237.8$ Kbps \cite{gong2023efficient}. Subcarrier-level designs further exploit the frequency-domain structure within each OFDM symbol, pushing the peak throughput to $743$ Kbps \cite{yu2023subscatter}. More recent sample-level approaches operate at an even finer granularity, manipulating shorter waveform segments and achieving peak throughputs of up to $10$ Mbps \cite{qin2024pushing}.

This throughput gain, however, comes with increasing implementation cost. Packet- and symbol-level modulation can remain compatible with commodity WiFi receivers because their effects survive standard WiFi demodulation and decoding. In contrast, subcarrier- and sample-level modulation requires access to finer PHY information that commodity interfaces typically do not expose, thus relying on dedicated receivers such as software-defined radios (SDRs). Finer granularity also demands tighter synchronization and faster tag switching. Therefore, increasing modulation granularity trades commodity compatibility and implementation simplicity for higher throughput.

\subsection{Commercial Compatibility}
Unlike conventional RFID systems that typically rely on dedicated readers, WiFi backscatter has the potential to reuse widely deployed WiFi infrastructure, thereby reducing the cost and complexity of deployment. This has motivated growing interest in commodity-compatible WiFi backscatter, where tag data can be recovered directly by standard WiFi devices without specialized receiving hardware.

HitchHike \cite{zhang2016hitchhike} is an early representative of this direction. It introduces codeword translation, in which a tag transforms an incoming WiFi codeword into another valid codeword that can still be decoded by commodity WiFi receivers. TiScatter \cite{du2023timespan} encodes tag data into the time intervals between WiFi codewords, while Chameleon \cite{yuan2024native} first demodulates ambient WiFi signals and then reshapes them into valid WiFi packets that embed tag information.

Despite this progress, seamless integration with existing WiFi infrastructure remains challenging. Most existing systems are still research prototypes and are typically evaluated under controlled WiFi transmissions or require specific assumptions about ambient WiFi traffic. These constraints limit their direct deployment in practical WiFi networks.

\subsection{Towards Simplified Deployment}
Codeword translation \cite{zhang2016hitchhike} is widely adopted in WiFi backscatter because it allows tag data to be embedded while preserving compatibility with standard WiFi decoding. However, recovering the tag data often requires knowledge of the original WiFi data as a reference, which has led many systems to rely on an additional receiver. This extra hardware increases deployment complexity and is difficult to accommodate on commodity devices, which typically provide only a single WiFi radio. These limitations have motivated the development of single-receiver WiFi backscatter systems.

RapidRider \cite{gong2023efficient} and ClusterFi \cite{wu2026clusterfi} embed reference information directly into WiFi packets, enabling tag data to be recovered from a single received stream. ParaFi \cite{wu2025concurrent} instead leverages the known pilot symbols in OFDM transmissions as an inherent reference for tag demodulation. Twoferscatter \cite{yang2023ambient} adopts a different strategy. Rather than explicitly acquiring the original WiFi data, the receiver enumerates candidate tag-bit values and reconstructs the corresponding candidate WiFi payloads. Because only the correctly reconstructed payload satisfies the packet frame check sequence (FCS), the receiver can identify the valid candidate and recover the tag data using a single WiFi receiver. These approaches reduce the reliance on dedicated reference hardware and move WiFi backscatter toward simpler and more practical deployment.

\subsection{Concurrent Communication}
As the number of tags in WiFi backscatter networks increases, sequential transmissions can become a major bottleneck to network throughput. To improve transmission efficiency, recent studies have explored concurrent backscatter communication, allowing multiple tags to transmit simultaneously while enabling the receiver to separate and recover their data.

Existing systems achieve concurrency through different signal-domain properties. ClusterFi \cite{wu2026clusterfi} exploits the clustering behavior of collided backscatter signals, where different combinations of tag bits produce distinguishable signal clusters that can be mapped back to the transmitted data. MultiRider \cite{wang2024multirider}, in contrast, exploits unaffected subcarriers in the frequency domain to recover tag data and employs successive interference cancellation to separate concurrent transmissions.

Despite their effectiveness, these approaches generally rely on fine-grained physical-layer information, such as raw in-phase and quadrature (IQ) samples or subcarrier-level measurements, which is typically unavailable from commodity WiFi devices. As a result, directly supporting concurrent backscatter communication on unmodified WiFi receivers remains challenging. This gap has motivated recent research toward commodity-compatible concurrent WiFi backscatter.

\subsection{Joint Communication and Sensing}
Early WiFi backscatter systems mainly focused on data communication, whereas recent work has begun to exploit backscattered signals for sensing and localization. This trend allows the same WiFi infrastructure to support both information delivery and environmental awareness.

LocTag \cite{wei2020loctag} leverages ambient WiFi signals from access points or smartphones together with WiFi-compatible backscatter modulation to localize commercial smartphones, achieving sub-meter accuracy in indoor experiments. TagFi \cite{soltanaghaei2021tagfi}, in contrast, focuses on localizing ultra-low-power WiFi tags using unmodified WiFi infrastructure. By modulating the tag reflection across multiple WiFi packets, TagFi distinguishes the backscatter path from other multipath components in the measured channel state information (CSI). It then estimates the angular information of the backscatter and line-of-sight paths and uses their geometric relationship to infer the tag location with a single WiFi receiver.

\section{Practical Challenges}

Despite substantial progress, WiFi backscatter still faces several challenges to widespread deployment that have received limited attention and are discussed in this section.

\subsection{Limited Transmitter-to-Tag Operating Range}
Signal detection is inherently more challenging in WiFi backscatter than in conventional RFID systems. An RFID tag is typically powered by a continuous reader signal and can modulate the carrier once activated. WiFi signals, however, are transmitted in bursts, with idle intervals between packets. A WiFi backscatter tag must therefore detect each incoming packet and determine its timing before performing modulation.

Most existing prototypes use passive envelope detectors or active RF detectors for packet detection. Their limited sensitivity, however, restricts the distance over which WiFi packets can be reliably detected. As the transmitter-to-tag distance increases, the incident signal strength decreases and may eventually fall below the detector sensitivity threshold, causing missed packet detections and unreliable backscatter operation. Consequently, the transmitter-to-tag range of many existing prototypes is limited to approximately one meter or less, constraining the deployment flexibility and practical applicability of WiFi backscatter systems.

\subsection{Limited System Generality}
Existing WiFi backscatter systems often rely on specific protocol features or receiver architectures, which limits their applicability across diverse WiFi environments. For example, HitchHike \cite{zhang2016hitchhike} enables commodity decoding but requires two receivers to separately capture the original and backscattered signals. CAB \cite{gong2025universal} supports joint recovery of ambient WiFi and tag data using a single receiver, but depends on SDR-based signal processing. Chameleon \cite{yuan2024native}, in contrast, enables direct communication with commodity WiFi devices but currently targets only 802.11b. These protocol- and architecture-specific constraints make it difficult for a single design to operate across heterogeneous WiFi platforms and standards.

\subsection{High Packet Loss in Frequency-Shifted Backscatter}
Frequency shifting \cite{zhang2016enabling} is commonly employed in WiFi backscatter to separate the reflected signal from the strong direct-link WiFi signal and thereby mitigate interference. However, shifting the backscattered signal to another channel can significantly degrade packet reception. In practice, we observe that only a portion of the frequency-shifted packets can be successfully decoded on the target channel, leading to reduced effective throughput and communication reliability. Despite its practical importance, this issue has received comparatively limited attention in existing WiFi backscatter research. A better understanding of the underlying causes and more robust reception mechanisms are therefore needed to improve the reliability and efficiency of frequency-shifted WiFi backscatter systems.

\subsection{Commodity-Compatible Concurrent WiFi
Backscatter}
Practical IoT deployments often involve multiple tags operating within the same environment. For instance, a smart home may contain numerous battery-free tags that periodically report temperature, motion, or health-related information. Due to their limited power and hardware constraints, however, backscatter tags are generally unable to perform conventional carrier sensing and channel coordination, making simultaneous transmissions likely in practical IoT deployments.

Recent studies have shown that concurrently backscattered WiFi signals can be separated and decoded by exploiting fine-grained physical-layer information \cite{wu2026clusterfi,wang2024multirider}. However, such information is typically unavailable from commodity WiFi devices, preventing these approaches from being directly implemented on unmodified receivers. Thus, existing concurrent backscatter systems require specialized receiving hardware, increasing deployment cost and limiting their ability to reuse existing WiFi infrastructure. Consequently, enabling concurrent multi-tag decoding on unmodified commodity WiFi receivers remains an important challenge for the practical adoption of WiFi backscatter.

\subsection{Lack of Standardization}
The widespread adoption of RFID has been facilitated by standardized protocols that define how readers discover, activate, identify, and communicate with tags. WiFi backscatter, however, still lacks a unified protocol framework. Existing systems generally adopt their own designs for signal excitation, tag modulation, channel access, and data decoding, resulting in limited interoperability across different tags and WiFi devices. Establishing common protocols and interfaces is therefore essential for integrating WiFi backscatter into existing wireless ecosystems and facilitating its practical deployment.

\section{Conclusion}

WiFi backscatter provides a promising foundation for low-power IoT connectivity by reusing widely available WiFi infrastructure. This article surveyed its fundamental principles, representative designs, emerging applications, and recent research trends, with particular attention to the challenges that affect real-world deployment. Although significant progress has been made, practical limitations remain in transmitter-to-tag operating range, packet reception reliability, system generality, commodity-compatible concurrent transmission, and standardization. Overcoming these limitations will be critical to enabling WiFi backscatter to move beyond research prototypes and become a practical communication technology for future green IoT systems.

\section{Acknowledgment}
This work received no external financial support. Large language models were used for proofreading and for the creation of visual elements in Figs. \ref{fig_overview} and \ref{fig_research_trends}, including illustrations of the WiFi transmitter, tag, receiver, waveforms, and application scenarios.

\section{Biographies}
\textbf{Weiqi Wu} (weiqiwu@ustc.edu.cn) is currently a postdoctoral researcher with the School of Computer Science and Technology, University of Science and Technology of China (USTC). His research interests include green IoT and backscatter communication.

\textbf{Shuo Wang} (iaurorawang@gmail.com) is currently an engineer with the CSSC Systems Engineering Research Institute. His research interests include sustainable IoT, health management, and cyber-physical systems.

\textbf{Zhaoyuan Xu} (zhaoyuxu@cityu.edu.hk) is currently a postdoctoral researcher with the Department of Computer Science, City University of Hong Kong. His research interests include passive IoT, backscatter communication, and wireless technologies.

\bibliographystyle{IEEEtran}
\bibliography{references}


\vfill

\end{document}